\documentclass[3p,12pt]{elsarticle}
\biboptions{sort&compress}
\usepackage{lineno,hyperref}
\modulolinenumbers[5]
\usepackage{amssymb,epsfig,bm,mathrsfs}
\usepackage[T1]{fontenc}
\usepackage[utf8]{inputenc}
\usepackage{times,txfonts}
\usepackage{isotope}
\usepackage{xcolor}
\usepackage{graphicx}
\usepackage{gensymb}
\usepackage{tikz}
\newcommand{\be}{\begin{equation}}
\newcommand{\ee}{\end{equation}}
\newcommand{\bea}{\begin{eqnarray}}
\newcommand{\eea}{\end{eqnarray}}
\newcommand{\de}{\partial}
\newcommand{\vectau}{{\bm \tau}}
\newcommand{\vecrho}{{\bm \rho}}
\newcommand{\aap}{{Astron. and Astrophysics}}
\newcommand{\nat}{{Nature}}
\newcommand{\prc}{{Phys. Rev. C}}
\newcommand\myeqref[1]{(\ref{#1})}
                                                                                      
\begin{document}
\begin{frontmatter}

\title{Constraining hypernuclear density functional with $\Lambda$-hypernuclei and compact stars}

\author[a]{E. N. E. van Dalen}
\ead{eric.van-dalen@uni-tuebingen.de}
\author[b]{Giuseppe Colucci}
\ead{colucci@th.physik.uni-frankfurt.de}
\author[b]{Armen Sedrakian}
\ead{sedrakian@th.physik.uni-frankfurt.de}

\address[a]{Institute for Theoretical Physics, T\"ubingen University, 
 D-72076 T\"ubingen, Germany }
\address[b]{Institute for Theoretical Physics, J.-W. Goethe University, 
 D-60438 Frankfurt am Main, Germany}

\begin{abstract}
  We present a simultaneous calculation of heavy single-$\Lambda$
  hypernuclei and compact stars containing hypernuclear core within a
  relativistic density functional theory based on a Lagrangian which
  includes the hyperon octet and lightest isoscalar-isovector mesons
  which couple to baryons with density-dependent couplings. The
  corresponding density functional allows for SU(6) symmetry breaking
  and mixing in the isoscalar sector, whereby the departures in the
  $\sigma$-$\Lambda$ and $\sigma$-$\Sigma$ couplings away from their
  values implied by the SU(3) symmetric model are used to adjust the
  theory to the laboratory and astronomical data.  We fix
  $\sigma$-$\Lambda$ coupling using the data on the single-$\Lambda$ hypernuclei
  and derive an upper bound on the $\sigma$-$\Sigma$ from the
  requirement that the lower bound on the maximum mass of a compact
  star is $2 M_{\odot}$.
\end{abstract}

\begin{keyword}
Neutron stars\sep Hypernuclei\sep Equations of state:nuclear matter\\
97.60.Jd\sep 21.80.+a\sep   21.65.Mn 
\end{keyword}

\end{frontmatter}

\section{Introduction}  The current and upcoming experimental studies 
of the properties of $\Lambda$-hypernuclei in laboratory,
such as HKS experiment at JLab in the US, J-PARC experiment in Japan,
PANDA experiment at FAIR in Germany, the  ALICE experiment at CERN, 
will greatly advance our understanding of the strange sector
of the nuclear forces and properties of hypernuclei. 
Astronomical motivation to study hypernuclear stars resurged after the
recent observations of two-solar-mass pulsars in binary orbits with
white dwarfs~\cite{2010Natur.467.1081D,2013Sci...340..448A}. Hyperons
become energetically favorable once the Fermi energy of neutrons
exceeds their rest mass. The onset of hyperons reduces the degeneracy
pressure of a cold thermodynamic ensemble, therefore, the equation of
state (EoS) becomes softer than in the absence of the hyperons.  As a result
the maximum possible mass of a compact star decreases to values which
contradict the observations. This contradiction is known as ``hyperonization puzzle''.   

What can be said about the effective amount of attraction of
hypernuclear forces? The experimental observations of bound
$\Lambda$-hypernuclei imply that the interaction must be attractive
enough to bind a $\Lambda$ particle to a medium and heavy mass nucleus.
At the same time the existence of two-solar-mass pulsars requires
sufficient repulsion (at least at high densities) to guarantee the stability of
hypernuclear compact stars, if such exist. Therefore, {\it the combined laboratory
  and astronomical data limit from above and below the attraction
  among hyperons in nuclear medium in any particular model.}

In this work we use a relativistic density functional theory  (DFT) of
hypernuclear matter to extract these bounds.  Density functional
theory is a very successful theoretical tool to study complex
many-body systems in various fields including strongly correlated
electronic systems, quantum chemistry, atomic and molecular systems,
classical liquids, magnetic materials, etc.~\cite{Engel2013}.  In
particular, relativistic covariant DFTs have been applied to study
bulk hypernuclear systems and compact stars both in the past (see, for
example, \cite{WeberBook,2007PrPNP..58..168S} for an account of the early work) and in
recent years, notably to address the ``hyperonization puzzle''
~\cite{Ryu2011,Lastowiecki2011,Weissenborn2012,
Bonanno2012,Bednarek2012,2012EL.....9739002M,Providencia2013,Chamel2013,2013PhRvC..87a5804D}.
Glendenning and Moszkowski~\cite{Glendenning1991} were the first to
recognize the importance of reconciliation of neutron-star masses and
binding energies of the $\Lambda$-hypernuclei. Since the recent
discoveries of heavy compact stars the astronomical constraints have
become much tighter. The quality of relativistic density functionals
have considerably improved in the last decade due to better
constraints from the phenomenology of nuclei~\cite{DDME2}.
Here we use an extension of nuclear density functional with a
density-dependent parameterization of the couplings~\cite{DDME2}, which
 was extended to the hypernuclear sector in
Ref.~\cite{Colucci2013} within the SU(3) symmetric model.  The focus
of that work was on the sensitivity of the EoS of hypernuclear matter
to the unknown hyperon–scalar-meson couplings. Within this framework,
it was argued that the parameters can be tuned such that two-solar
mass  hyperonic compact stars emerge (which is not possible within
the standard SU(3) parameterization).

Here we test this model by carrying out calculations of a number of
hypernuclei and by providing a {\it combined constraint} on the
parameters of the underlying DFT by invoking both the astronomical and
laboratory data on hypernuclear systems. We show that the coupling of
$\sigma$-meson to the $\Lambda$-hyperon can be optimized to fit the
data on hypernuclei, thus narrowing down the parameter space. We then
constrain the parameter space of the remaining $\sigma-\Sigma$
coupling using some general inequalities as well as the astronomical
observations of  $2M_{\odot}$ pulsars.

\section{Density functional theory of hypernuclear matter} The
relativistic Lagrangian density of our model reads
\bea\label{eq:lagrangian} \nonumber {\cal L} & = &
\sum_B\bar\psi_B\bigg[\gamma^\mu \left(i\de_\mu-g_{\omega B}\omega_\mu
  - \frac{1}{2} g_{\rho B}\vectau\cdot\vecrho_\mu\right)
- (m_B - g_{\sigma B}\sigma)\bigg]\psi_B \\
\nonumber & + & \frac{1}{2}
\de^\mu\sigma\de_\mu\sigma-\frac{m_\sigma^2}{2} \sigma^2 -
\frac{1}{4}\omega^{\mu\nu}\omega_{\mu\nu} + \frac{m_\omega^2}{2}
\omega^\mu\omega_\mu - \frac{1}{4}\vecrho^{\mu\nu}\vecrho_{\mu\nu}
\\
& + & \frac{m_\rho^2}{2} \vecrho^\mu\cdot\vecrho_\mu
+\sum_{\lambda}\bar\psi_\lambda(i\gamma^\mu\de_\mu -
m_\lambda)\psi_\lambda,
\eea where the $B$-sum is over the $J^P = \frac{1}{2}^+$ baryon octet,
$\psi_B$ are the baryonic Dirac fields with masses $m_B$.  The meson
fields $\sigma,\omega_\mu$ and $\vecrho_\mu$ mediate the interaction
among baryon fields, $\omega_{\mu\nu}$ and $\vecrho_{\mu\nu}$
represent the field strength tensors of vector mesons and
$m_{\sigma}$, $m_{\omega}$, and $m_{\rho}$ are their masses. The
baryon-meson coupling constants are denoted by $g_{mB}$. The last line
of Eq.~(\ref{eq:lagrangian}) stands for the contribution of the free
leptons, where the $\lambda$-sum runs over the leptons
$e^-,\mu^-,\nu_e$ and $\nu_\mu$ with masses $m_\lambda$. The density
dependence of the couplings implicitly takes into account many-body
correlations among nucleons which are beyond the mean-field
approximation.  The nucleon-meson coupling constants are parametrized
as $g_{iN}(\rho_B) = g_{iN}(\rho_{0})h_i(x)$, for $i=\sigma,\omega$,
and $g_{\rho N}(\rho_B) = g_{\rho N}(\rho_{0})\exp[-a_\rho(x-1)]$ for
the $\vecrho_\mu$-meson, where $\rho_B$ is the baryon density,
$\rho_0$ is the saturation density, $x = \rho_B/\rho_0$ and the
explicit form of the functions $h_i(x)$ and the values of couplings
can be found elsewhere~\cite{DDME2,Colucci2013}. This density
functional is consistent with the following parameters of nuclear
systems: saturation density $\rho_0=0.152$ fm$^{-3}$, binding energy
per nucleon $E/A=-16.14$ MeV, incompressibility $K_0=250.90$ MeV,
symmetry energy $J=32.30$ MeV, symmetry energy slope $L=51.24$ MeV,
and symmetry incompressibility $K_{sym} = -87.19$ MeV all taken at
saturation density~\cite{2011PhRvC..83d5810D}.  These values of
parameters are in an excellent agreement with the nuclear
phenomenology~\cite{Experiment}. The third order derivatives
of the energy and symmetry energy with respect to density taken at
saturation have the following values $Q_0=478.30$ and $Q_{sym}=
777.10$ MeV.

The pressure and
energy density of the model is further supplemented by the
contribution coming from the so-called rearrangement
self-energy~\cite{1995PhRvC..52.3043F,1999NuPhA.656..331T}, which guarantees the
thermodynamical consistency. The hyperon--meson couplings are fixed
according to the SU(3)-flavor symmetric octet model.  Due to the
universal coupling of the $\vecrho_\mu$ meson to the isospin current
and the ideal mixing between the $\omega$ and $\phi$
mesons~\cite{2009JHEP...07..105K}, the couplings between hyperons and
vector mesons are as follows:
\be \label{vector_couplings} x_{\rho \Xi}
= 1, \quad x_{\rho \Sigma}= 2, \quad x_{\omega \Xi} = \frac{1}{3},\quad
x_{\omega \Sigma} = x_{\omega \Lambda} = \frac{2}{3}, \quad
x_{\rho \Lambda} = 0.  
\ee 
where $x_{\rho \Xi} = g_{\rho \Xi}/g_{\rho N}$,
$x_{\rho \Sigma} = g_{\rho \Sigma}/ g_{\rho N}$, etc.

Within the octet model the baryon--scalar-mesons couplings of the
scalar octet can be expressed in terms of only two parameters, the
nucleon--$a_0$-meson coupling constant $g_S$ and the $F/(F+D)$ ratio
of the scalar octet~\cite{DeSwart1963}. Allowing for mixing of the 
scalar singlet state, the couplings of the baryons with the
$\sigma$-meson obey the following relation~\cite{Colucci2013}: $
2(g_{\sigma N} + g_{\sigma\Xi}) = 3g_{\sigma \Lambda} +
g_{\sigma \Sigma}$. We assume that the hyperon coupling constants must
be positive and less than the nucleon coupling constants.  Solving this
equation for one of the dependent hyperon--$\sigma$-meson coupling
constant, say $g_{\sigma \Xi}$, one finds
\be \label{additional_constraints2} 1\le \frac{1}{2} (3
x_{\sigma \Lambda} + x_{\sigma \Sigma}) \le 2.  
\ee 
These inequalities define a bound on the area spanned by the coupling
constants $x_{\sigma \Lambda}$ and $x_{\sigma \Sigma}$, which we will
constrain further in the following.

\section{Finite nuclei}  We now apply the same density functional, which
is derived from the Lagrangian (\ref{eq:lagrangian}) to finite
$\Lambda$-nuclei. For alternative applications of relativistic density
functionals to finite $\Lambda$-hypernuclei see, for example,
Refs.~\cite{1990PhRvC..42.2469R,2000PhRvC..61f4309K,2009NuPhA.831..163F}
and references therein.
The Hamiltonian for protons, neutrons, and $\Lambda$-hyperons is the
sum of the nuclear Hartree-Fock (HF) part and the Coulomb contribution
which acts only among the charged particles (here protons)
\begin{equation}
H_{\rm HF,B}=H_{\rm RMF,B} + H_{\rm Coul} \delta_{B p},
\label{eq:HFHamil}
\end{equation}
where $H_{\rm RMF,B}$ is the mean-field Hamiltonian corresponding to the
density functional discussed above and $H_{\rm Coul} $ denotes the Coulomb
contribution.  Note that the Hamiltonian is local. It is defined in
terms of the single-particle densities resulting from the eigenstates
of $H_{\rm HF,B}$, which implies that they have to be determined in a
self-consistent way.
The HF Hamiltonian is expressed in terms of the matrix elements
between the basis states $\langle \alpha \vert H_{\rm HF,B} \vert \beta \rangle$ of an
appropriate basis. The HF single-particle states $\vert \Psi_n \rangle$
are defined in terms of the expansion coefficients in this basis
\begin{equation}
\vert \Psi_n \rangle = \sum_{\alpha} \vert \alpha \rangle \langle \alpha \vert\Psi_n \rangle = 
\sum_{\alpha}
c_{n\alpha} \vert \alpha \rangle\,. 
\end{equation}
If the HF variational procedure is constrained to a spherical
description of hypernuclei, an appropriate basis system is formed by
spherical plane wave basis~\cite{Montani2004,vanDalen2009}, where a
baryon is freely moving in a spherical cavity with a radius $R$. Since
the HF Hamiltonian is already diagonal in the angular momentum quantum
numbers $j,l$, and $m$, it only needs to be diagonalized in the radial
quantum number.  The radial part of the wave function is then expanded
in terms of spherical Bessel functions.  The radius $R$ is chosen to
be large enough to guarantee that the results for the bound
single-particle states are insensitive to the changes in the value of
$R$. Furthermore, we choose the number of basis states high enough to
guarantee that the results are not affected by the truncation. The
eigenvalues, i.e. the single-particle energies, and eigenvectors,
i.e. the expansion coefficients, are then determined by matrix
diagonalizations of the Hamiltonians for the protons, neutrons, and
$\Lambda$-hyperons.

The total energy of a $\Lambda$-hypernucleus $E_{tot}$ can then be
obtained from the expression,
\begin{eqnarray}
E_{tot} = \frac{1}{2} \sum_{\alpha, B} \eta^B_\alpha ( t^B_\alpha + \varepsilon^B_\alpha)  + E_{rear} + E_{cm},
\label{eq:etot}
\end{eqnarray} 
where $\varepsilon^B_\alpha$ is the single-particle energy of the
$B$-baryon, $t^B_\alpha$ is its kinetic energy, and $\eta^B_\alpha$ is
its occupation factor. Because the couplings of our model are
density-dependent we need to include the rearrangement contribution
$E_{rear}$ to insure the consistency of the model.  Finally, the center of mass
correction is given by $  E_{cm} = - ({1}/{2M})\langle \bold{P}_{cm}^2 \rangle $,
with
\begin{equation}
  \langle \bold{P}_{cm}^2 \rangle
     =  \sum_\alpha \eta_\alpha \langle \alpha | \bold{p}_\alpha^2 | \alpha \rangle
         - \sum_{\alpha\beta}  ( \eta_\alpha \eta_\beta + \zeta_\alpha \zeta_\beta)
	     \, \langle \beta | \bold{p}_\alpha | \alpha \rangle
	      \cdot \langle \alpha | \bold{p}_\beta | \beta \rangle,\nonumber
\end{equation}
where $\zeta_\alpha$ is the anomalous occupation factor.  To explore
the sensitivity of the results on the coupling of $\Lambda$-hyperon to
mesons we consider three sets of parameters:  model $a$ with 
$x_{\sigma\Lambda} = 0.52$, model $b$ with $x_{\sigma\Lambda} = 0.59$,
and model $c$ with $x_{\sigma\Lambda} = 0.66$; all three models have
$x_{\omega\Lambda} = 2/3$ and $x_{\rho\Lambda}= 0$.  The HF
calculations were carried out in a spherical box with a radius of $R=15$
 fm using the spherical plane wave basis for the
$\Lambda$-hypernuclei $^{17}_{\Lambda}$O, $\isotope[41][\Lambda]{Ca}$,
and $^{49}_{\Lambda}$Ca.  The results of these calculations are
presented in Table~\ref{table:1}, where we list the single-particle
energy of the $\Lambda$ 1s$_{1/2}$ state, the binding energy of the
nucleus and the rms radii $r_B$ for neutrons, protons and the 
$\Lambda$-hyperon.  
\begin{table}
\begin{center}
\begin{tabular}{|c|c|c|ccc|}
\hline
& \qquad $\Lambda$ 1s$_{1/2}$ state\qquad &\qquad $E/A$ &\qquad  $r_{p}$\qquad  & \qquad $r_{n}$ \qquad  & \qquad  $r_{\Lambda}$ \qquad\\
&\qquad  [MeV] \qquad & \qquad [MeV] \qquad & \qquad [fm] \qquad  &\qquad  [fm] \qquad  & \qquad  [fm]\qquad     	\\
\hline
{$\isotope[17][\Lambda]{O}$}  & & & & &  	\\
\hline
$a$  & 0.846 &  $-7.443$    &  2.609   &  2.579  &  8.313   \\
$b$& $-4.564$ &  $-7.760$   &  2.606   &  2.576   &  3.203   \\
$c$ & $-27.279$  &  $-9.035$    & 2.563    & 2.534   &  1.977   \\
\hline
\hline
{$\isotope[41][\Lambda]{C}$ } & & & & &  	\\
\hline
$a$& 0.934  &  $-8.336$     & 3.372     &  3.319  &  8.710   \\
$b$& $-8.519$ &  $-8.565 $  &  3.370   &  3.317   &  3.168   \\
$c$& $-35.224$ &  $-9.199$     &  3.347    & 3.294    &  2.298  \\
\hline
\hline
{$\isotope[49][\Lambda]{C}$}& & & & & 	\\
\hline
$a$ & 0.973 &  $-8.442$    &  3.389    & 3.576   &  8.825   \\
$b$ & $-9.882$ &  $-8.662$    &  3.387    & 3.571     &  3.140   \\
$c$& $-37.257$ & $-9.207$     & 3.365     &  3.548  &  2.419   \\
\hline
\end{tabular}
\caption{
\label{table:1} 
Properties of $\Lambda$-hypernuclei $\isotope[17][\Lambda]{O}$,
$\isotope[41][\Lambda]{C}$, and $\isotope[49][\Lambda]{C}$ for the
models $a$, $b$, and $c$. The columns list the single-particle energy
of the $\Lambda$ 1s$_{1/2}$ state, the binding energy and the rms
radii for neutrons, protons and $\Lambda$-hyperon.  }
\end{center}
\end{table}
\begin{figure}[!]
\begin{center}
\includegraphics[width=0.6\columnwidth] {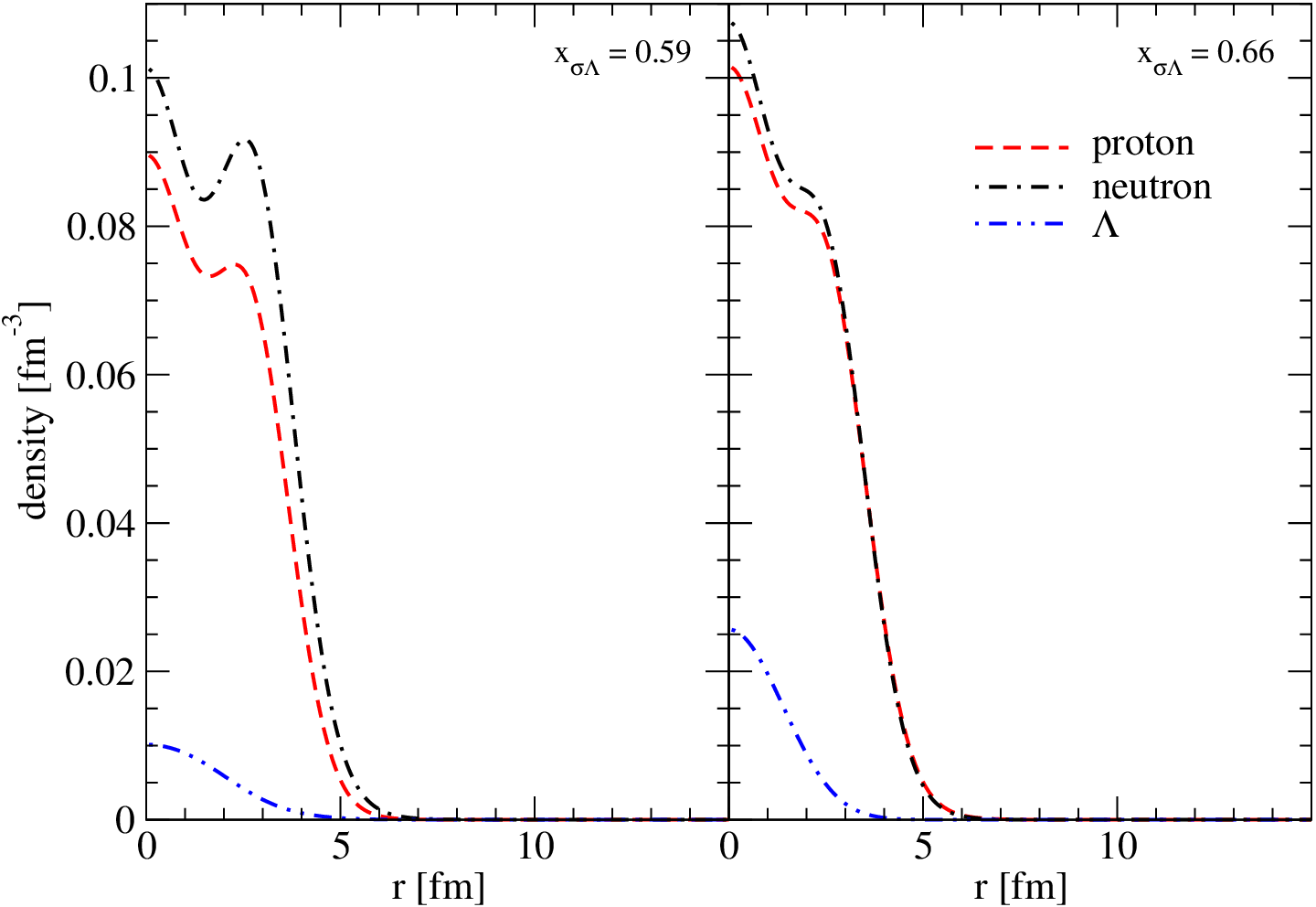}
\caption{(Color online.) Proton (dashed), neutron (dashed-dotted), and
 $\Lambda$ (dashed-double dotted) density distributions in
$\isotope[49][\Lambda]{C}$ for the models $b$ (left panel) and $c$
(right panel).}
\label{fig:dendisCa49Lambda}
\end{center}
\end{figure}
\begin{table}
\begin{center}
\begin{tabular}{|c|c|c|c|ccc|}
\hline
& $E_{Mass}[\Lambda$ 1s$_{1/2}]$ 
& $E[\Lambda$ 1s$_{1/2}$]  & $E/A$  &
$r_{p}$  & $r_{n}$ &  $r_{\Lambda}$    \\
&  [MeV]    &  [MeV]   &  [MeV] & [fm]  &[fm]  &  [fm]           \\
\hline
\isotope[17][\Lambda]{O} \ \ \ &  $-12.109$  &    $-11.716 $ &   $-8.168 $  &  2.592    &  2.562  &   2.458  \\
$\isotope[16]{O}$ & $-$ &$ -$ &  $-8.001$    & 2.609     & 2.579   &  $-$   \\
\hline
$\isotope[41][\Lambda]{C}$ 
&  $-17.930$ &   $-17.821$ &   $-8.788$   &  3.362    & 3.309  &  2.652   \\
$\isotope[40]{Ca}$ & $-$ & $-$ & $-8.573$     & 3.372     &  3.320  &  $-$  \\
\hline
$\isotope[49][\Lambda]{Ca}$ 
& $-19.215$  &  $-19.618$ & $-8.858$ &  3.379    & 3.562 & 2.715  \\
$\isotope[48]{Ca}$ & $-$ & $-$ & $-8.641$ &  3.389    & 3.576 &  $-$  \\
\hline
\end{tabular}
\end{center}
\caption{\label{table:M4} Single-particle energies 
of the $\Lambda$ 1s$_{1/2}$ states, binding energies, and rms radii 
of the $\Lambda$-hyperon, neutron, and proton of 
\isotope[17][\Lambda]{O}, $\isotope[41][\Lambda]{C}$, and $^{49}_{\Lambda}$Ca are 
presented for optimal model. In addition, single-particle energies of 
the $\Lambda$ 1s$_{1/2}$ states, i.e. separation energies of 
the $\Lambda$-particle,  obtained from the mass formula of 
Ref.~\cite{Levai1998} are given for these $\Lambda$-hypernuclei.
Furthermore, the properties of $^{16}$O, $^{40}$Ca, and $^{48}$Ca 
are given for the optimal model.}
\end{table}
Model $a$ with the smallest value of the $\Lambda$-$\sigma$
coupling predicts a positive single-particle energy for the $\Lambda$
1s$_{1/2}$ state of these nuclei, which means that the 
$\Lambda$-hyperon is not bound. The fact that one does not have a hypernucleus
is also reflected in the unusually large value of $r_{\Lambda}$.  The other
two models with larger values of the $\Lambda$-$\sigma$ coupling
predict negative single-particle energies of the $\Lambda$ 1s$_{1/2}$
state.  It is seen from Table~\ref{table:1} that a larger value of the
$\Lambda$-$\sigma$ coupling yields a larger binding energy and a
smaller $r_{\Lambda}$.
\begin{figure}[h]
\begin{center}
 \includegraphics[width=0.52\columnwidth]{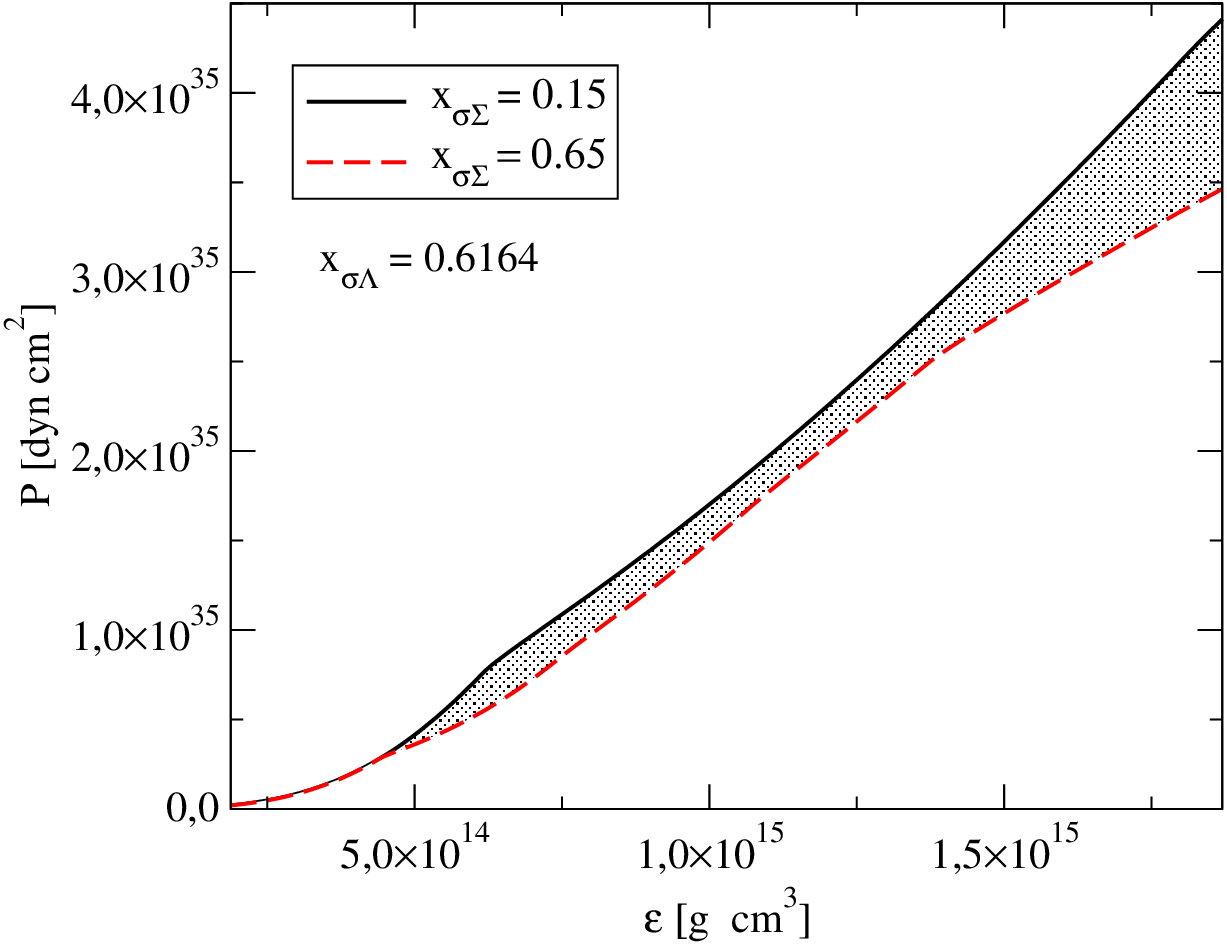}
\end{center}
   \vskip-2ex
 \caption{\label{fig:EoS_hypernuclear} (Color online.)
Zero temperature equations of state of hypernuclear matter for fixed  
$x_{\sigma\Lambda} = 0.6164$ and a range of values 
$ 0.15\le x_{\sigma\Sigma}\le 0.65 $. These values generate the shaded
area, which is bound from below by the softest EoS
(dashed red line) corresponding to $ x_{\sigma\Sigma}= 0.65 $
and from above by the hardest EoS (solid line) corresponding to 
$ x_{\sigma\Sigma}= 0.15 $.
}
\end{figure}

In Fig.~\ref{fig:dendisCa49Lambda} we show the proton, neutron, and $\Lambda$
density distributions in $^{49}_{\Lambda}$Ca. Since model
$a$ with its smallest value of the $\Lambda$-$\sigma$ coupling clearly
contradicts experimental data, only models $b$ and $c$ are
considered. The model $c$ which has  the largest value of the
$\Lambda$-$\sigma$ coupling  predicts the highest central
$\Lambda$ density for $^{49}_{\Lambda}$Ca. Also the neutron and proton
density distributions are to some extent affected by the large value
of the $\Lambda$-$\sigma$ coupling.

Table~\ref{table:1} shows clearly that the single-particle energy of
the $\Lambda$ 1s$_{1/2}$ state is very sensitive to the value of the
$\Lambda$-$\sigma$ coupling. The experimental data on properties of a
number of $\Lambda$-hypernuclei, such as the single-particle energy of
the $\Lambda$ 1s$_{1/2}$ state, has been used to construct a mass
formula~\cite{Levai1998}, which extends the familiar
Bethe-Weizs\"acker mass formula to include in addition to the
non-strange nuclei the $\Lambda$-hypernuclei. A comparison with the
predictions of this mass formula shows that the $\Lambda$ 1s$_{1/2}$
states in the model $b$ are too weakly bound, whereas those in the
model $c$ are too strongly bound. Therefore, we proceed further to
fine-tune the $x_{\sigma \Lambda}$ coupling in order to fit the values
of the single-particle energies, i.e. separation energies of the
$\Lambda$ particle, obtained from the mass formula of
Ref.~\cite{Levai1998}. The {\it optimal model} obtained in this way
has $x_{\sigma \Lambda}=0.6164$.  Within this optimal model we have
recomputed the properties of $^{17}_{\Lambda}$O,
$\isotope[41][\Lambda]{Ca}$, and $\isotope[49][\Lambda]{Ca}$.  The
results are given in Table~\ref{table:M4}, where we observe that the
rms radius of the $\Lambda$ in the 1s$_{1/2}$ state increases with
increasing mass number, which is in agreement with other theoretical
models~\cite{1990PhRvC..42.2469R,2000PhRvC..61f4309K,2009NuPhA.831..163F}.

In Table~\ref{table:M4}, the properties of $^{16}$O, $^{40}$Ca, and
$^{48}$Ca are also given to investigate the effects of the $\Lambda$
hyperon on the nucleons. The binding energies of $^{17}_{\Lambda}$O,
$\isotope[41][\Lambda]{Ca}$, and $\isotope[49][\Lambda]{Ca}$ are
larger than those of $\isotope[16]{O}$, $\isotope[40]{Ca}$, and
$\isotope[48]{Ca}$, respectively.  This can be explained by the fact
that the nucleon single-particle states are slightly deeper due to the
presence of the $\Lambda$-hyperon. In addition, the rms radii of the
nucleons are slightly smaller (by about 0.01 to 0.02 fm) in the
$\Lambda$-hypernuclei.  However, the addition of the $\Lambda$-hyperon
to $\isotope[16]{O}$ and $\isotope[40]{Ca}$ does not change the
differences between neutron and proton radii (neutron skin). The
values of neutron skin in these nuclei are $r_{n}-r_{p} =
-0.030$ and $-0.052$ fm, respectively. Only in the case of $\isotope[48]{Ca}$
we observe a small change: the neutron skin changes  from 0.187 fm 
in $\isotope[48]{Ca}$ to 0.184 fm in $\isotope[49][\Lambda]{Ca}$.

\section{Compact stars} The recent observations of two-solar-mass
pulsars in binary orbits with white
dwarfs~\cite{2010Natur.467.1081D,2013Sci...340..448A} place an
observational lower bound on the maximum mass of any sequence of
compact stars based on the unique equation of state (hereafter EoS) of
dense matter.  Massive compact stars may demand substantial population
of heavy baryons (hyperons). In Ref. \cite{Colucci2013} {\it both}
parameters $x_{\sigma\Lambda}$ and $x_{\sigma\Lambda}$ were varied
around the Nijmegen Soft Core (NSC) potential value $x_{\sigma
  \Lambda} = 0.58$ and $x_{\sigma \Sigma} =
0.448$~\cite{2006PhRvC..73d4009E} in a range that is consistent with
Eq.~\myeqref{additional_constraints2}.  In this section we revisit
this problem using additional insight gained from the studies of the
$\Lambda$-hypernuclei above. Specifically, we use the optimal model
from previous section to fix the value $x_{\sigma\Lambda} =
0.6164$. Then, we are left with the coupling $g_{\sigma\Sigma}$ which
is allowed to vary in the limits provided by
Eq. (\ref{additional_constraints2}).

The dependence of the EoS on the variation of the $\Sigma$-$\sigma$
coupling at $T=0$ at fixed value of $\Lambda$-$\sigma$
is shown in Fig.~\ref{fig:EoS_hypernuclear}.  The stiffest EoS is
obtained for the smallest value of $x_{\sigma \Sigma} = 0.15$. The
EoS band generated by the values of   $0.15 \le x_{\sigma \Sigma} \le
0.65$ is bounded from below by EoS which, as we shall see, is
incompatible with the lower bound on the maximum mass of a compact
star. Therefore, the parameter space included in this figure can be
narrowed down further by exploring the masses of corresponding
configurations. 
\begin{figure}[tb]
 \centering
\includegraphics[width=0.52 \columnwidth]{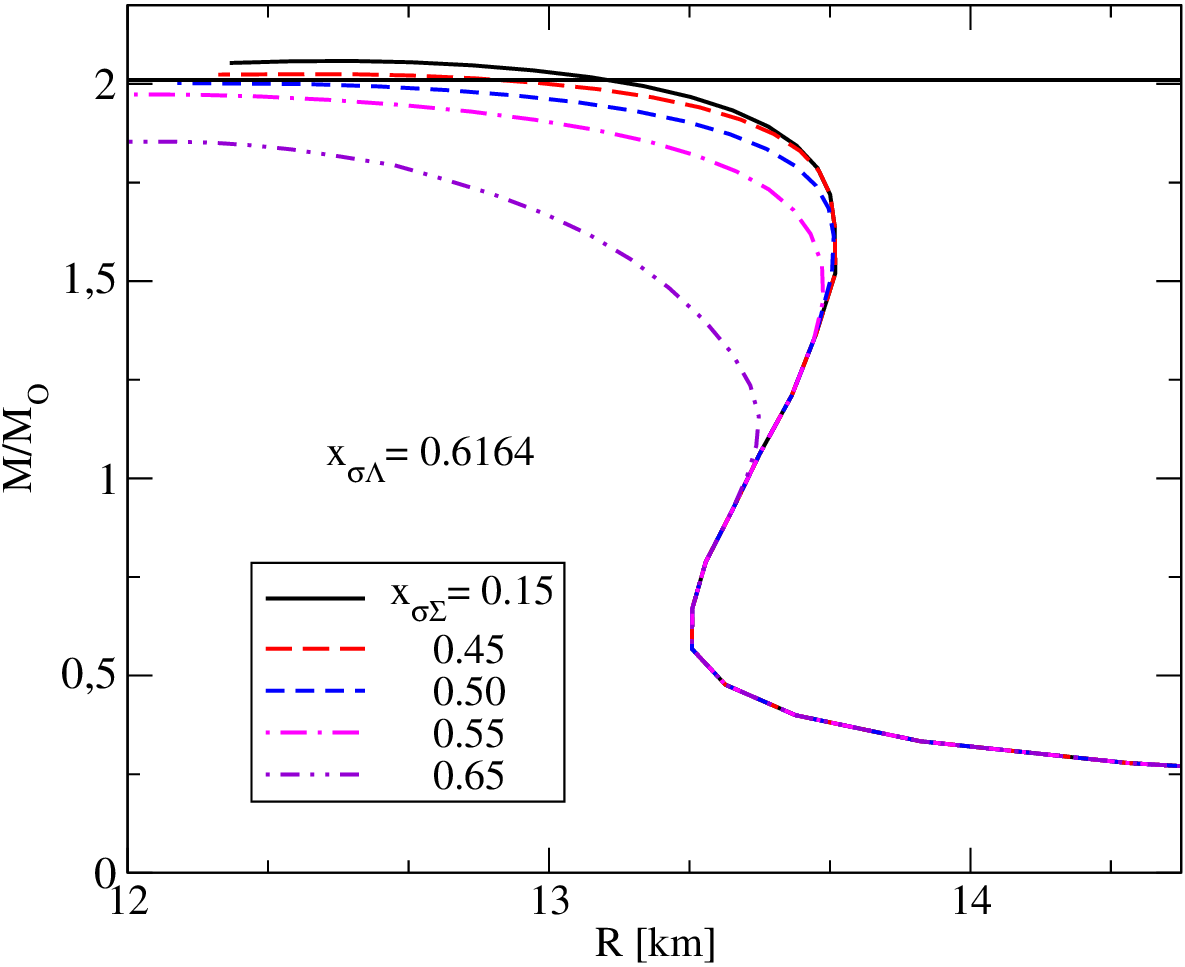}
\caption{ The mass--radius relations for compact hypernuclear stars at
  zero temperature. We fix $x_{\sigma\Lambda} = 0.6164$ and assign
  values to $ x_{\sigma\Sigma}$ from the range $ 0.15\le
  x_{\sigma\Sigma}\le 0.65 $ as indicated in the plot.  The horizontal
  line shows the observational lower limit on the maximum mass $2.01
  (\pm 0.04) M_{\odot}$~\cite{2013Sci...340..448A}.  }
 \label{fig:mass_vs_radius}
\end{figure}
\begin{figure}[!]
\centering
\vspace{0.4cm}
\includegraphics[width=0.52\columnwidth]{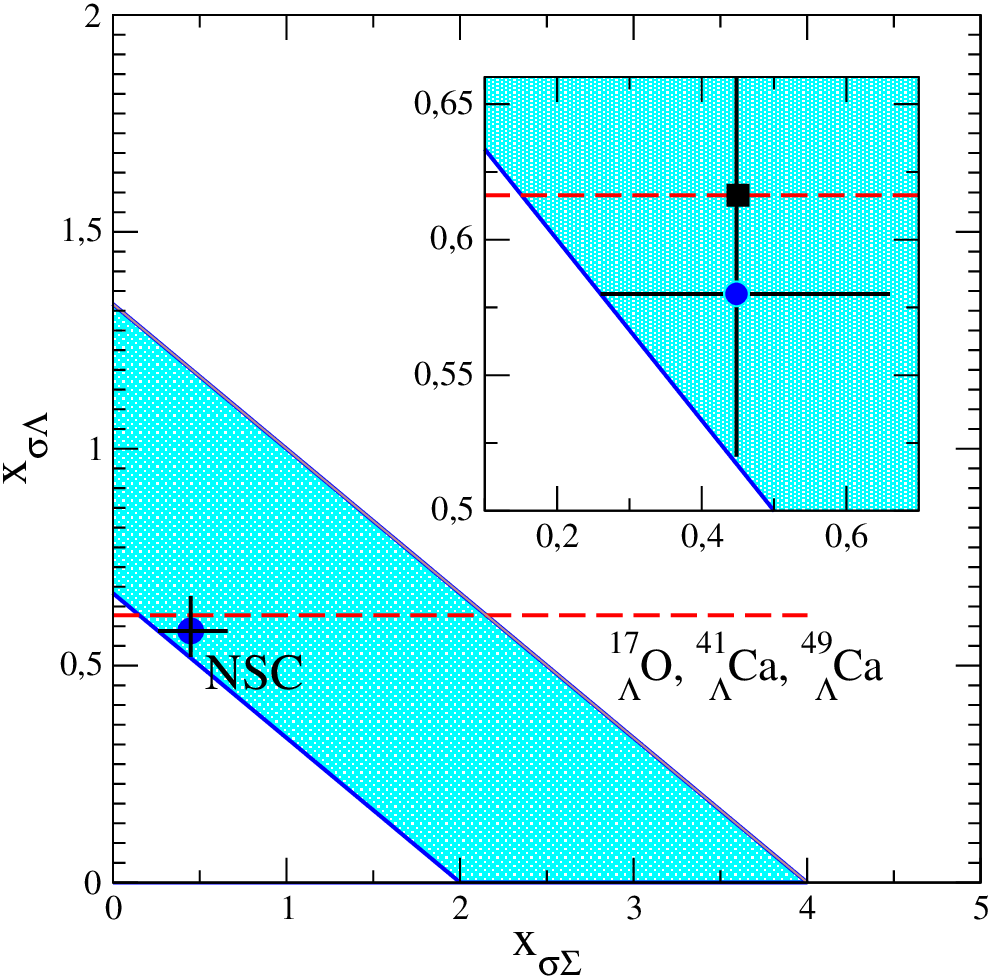}
\caption{(Color online.) The parameter space spanned by $x_{\sigma\Lambda}$ and
  $x_{\sigma\Sigma}$, where the inset enlarges the physically relevant
  area.  The shaded (blue online) area corresponds to the inequality
  \myeqref{additional_constraints2}. The dot corresponds to the values
  $x_{\sigma\Lambda} = 0.58$ and $x_{\sigma\Sigma} = 0.448$ derived
  from the Nijmegen Soft Core (NSC) potential. The dashed (red online)
  line shows the best fit value of $x_{\sigma\Lambda} = 0.6164$
  derived from hypernuclei. The square in the inset shows the limiting
  value of $x_{\sigma\Sigma} = 0.45$ for fixed $x_{\sigma\Lambda} =
  0.6164$ beyond which no stars with 2$M_{\odot}$ exist. }
\label{fig:hyperon_couplings}
\end{figure}
Fig.~\ref{fig:mass_vs_radius} shows the gravitational masses (in solar
units) vs radii for our sequences of stars. First, we see that large
enough masses can be obtained within the parameter range
covered. However, for large enough $x_{\sigma\Sigma}$ the maximum
masses of the sequences drop below the observational value
2$M_{\odot}$, specifically for $x_{\sigma\Lambda} = 0.6164$ this
occurs for $x_{\sigma\Sigma} \ge 0.45$.  The predicted radii of massive
hypernuclear stars are in the range of 13 km and are typically larger
than the radii of their purely nucleonic counterparts.

Fig.~\ref{fig:hyperon_couplings} shows the parameter space covered by
the coupling constants $x_{\sigma \Sigma}$ and $x_{\sigma
  \Lambda}$. The shaded (blue online) area is the parameter space
consistent with Eq.~\myeqref{additional_constraints2}. The dot
corresponds to the values of these parameters predicted by the
Nijmegen Soft Core (NSC) potential.  The dashed (red online) line
shows the optimal value of $x_{\sigma\Lambda}$ implied by the
hypernuclear data. The solid vertical and horizontal lines show the
parameter space explored in Ref.~\cite{Colucci2013}.  Finally, the
square in the inset shows the maximal value of $x_{\sigma\Sigma}\simeq
0.45$ (at fixed $x_{\sigma\Lambda}$) which is still consistent with the
2$M_{\odot}$ maximum value of a configuration. Thus, we conclude that
the optimal values of the parameters correspond to
\be\label{eq:optimal_x}
x_{\sigma\Lambda} = 0.6164, \qquad 0.15\le x_{\sigma \Sigma} \le 0.45.
\ee 
The first value is set by the study of (heavy) hypernuclei, the upper
limit of the second value is set by the 2$M_{\odot}$ constraint,
whereas the lower limit is set by the requirement of the consistency
with inequality \myeqref{additional_constraints2}.

\section{Conclusions} In this work we used a relativistic density
functional theory of hypernuclear matter to extract bounds on the
density-dependent couplings of a hypernuclear DFT. To do so, we used
simultaneous fits to the medium-heavy $\Lambda$-hypernuclei and the
requirement that the maximum mass of a hyperonic compact star is at
least two-solar masses.  This allowed us to narrow down significantly
the parameter space of couplings of DFT - the range of optimal values
of parameters is given in Eq.~\myeqref{eq:optimal_x}. Although our
work was carried out within a specific parameterization of the
hypernuclear density functional, it provides a proof-of-principle of
the method for constraining any theoretical framework that describes
hypernuclear systems using current laboratory and astrophysical data.

\section*{Acknowledgments} We are grateful to I. Mishustin,
H. M\"uther, C. Providencia, L. Rezzolla, D. Rischke, and J. Schaffner-Bielich for
discussions.  This work was supported by a grant (Mu 705/7-1) of the
Deutsche Forschungsgemeinschaft (E.N.E. v.D.), the HGS-HIRe graduate
program at Frankfurt University (G.C.) and by ``NewCompStar'', COST
Action MP1304.


\end{document}